\documentstyle[preprint,tighten,aps,epsfig]{revtex}

\begin{document}
\draft


\title {Generalization of the Peres criterion for local realism through nonextensive entropy}

\author{  Constantino Tsallis  $^{(a)}$  }

\address{Centro Brasileiro de Pesquisas Fisicas, 
Xavier Sigaud 150, 22290-180, 
Rio de Janeiro-RJ, Brazil,\\
and\\
Department of Mechanical Engineering, Massachusetts Institute of Technology, Rm. 3-164,
Cambridge, Massachusetts 02139, USA\\}

\author{Seth Lloyd $^{(b)}$ }

\address{Department of Mechanical Engineering, Massachusetts Institute of Technology, Rm. 3-160,
Cambridge, Massachusetts 02139, USA\\ }

\author{Michel Baranger$^{(c)}$}

\address{Center for Theoretical Physics, Laboratory for Nuclear Sciences and
Department of Physics,\\
Massachusetts Institute of Technology, 
Cambridge, Massachusetts 02139, USA\\}

\date{\today}
\maketitle


\begin{abstract}

A bipartite spin-1/2 system having the probabilities $\frac{1+3x}{4}$ of being in the Einstein-Podolsky-Rosen entangled state $|\Psi^-$$> \equiv \frac{1}{\sqrt 2}(|$$\uparrow>_A|$$\downarrow>_B$$-|$$\downarrow>_A|$$\uparrow>_B)$ and $\frac{3(1-x)}{4}$ of being orthogonal, is known to admit a local realistic description if and only if $x<1/3$ (Peres criterion). We consider here a more general case where the probabilities of being in the entangled states $|\Phi^{\pm}$$> \equiv \frac{1}{\sqrt 2}(|$$\uparrow>_A|$$\uparrow>_B \pm |$$\downarrow>_A|$$\downarrow>_B)$ and $|\Psi^{\pm}$$> \equiv \frac{1}{\sqrt 2}(|$$\uparrow>_A|$$\downarrow>_B \pm |$$\downarrow>_A|$$\uparrow>_B)$ (Bell basis) are given respectively by $\frac{1-x}{4}$, $\frac{1-y}{4}$, $\frac{1-z}{4}$ and $\frac{1+x+y+z}{4}$. Following Abe and Rajagopal, we use the nonextensive entropic form $S_q \equiv \frac{1- Tr \rho^q}{q-1}\;(q \in \cal{R}; \; $$S_1$$= -$ $Tr$ $ \rho \ln \rho)$ which has enabled a current generalization of Boltzmann-Gibbs statistical mechanics, and determine the entire region in the $(x,y,z)$ space where local realism is admissible. For instance, in the vicinity of the EPR state, classical realism is possible 
 if and only if $x+y+z<1$, which recovers Peres' criterion when $x=y=z$. In the vicinity of the other three states of the Bell basis, the situation is identical. A critical-phenomenon-like scenario emerges. These results illustrate the computational power of this new nonextensive-quantum-information procedure.

\end{abstract}


\bigskip

Quantum entanglement is a manifestation of the essential nonlocality of the quantum world, and a most intriguing physical phenomenon. It was first discussed as early as 1935 by Einstein, Podolsky and Rosen (EPR) \cite{EPR} and by Schroedinger \cite{schroedinger}, and has regained intensive interest in recent years due to its remarkable applications in quantum computation, teleportation and cryptography \cite{ekert,bennett,barenco,zurek,lloyd}, among others, as well as its connections to quantum chaos \cite{quantumchaos}.

Two systems $A$ and $B$ are said to be {\it uncorrelated} if and only if the density operator $\rho_{A+B}$ can be written as
\begin{equation}
\rho_{A+B} = \rho_A \otimes \rho_B \;\;\;\; (Tr_{A+B}\; \rho_{A+B}= Tr_A \;\rho_A=Tr_B \;\rho_B=1),
\end{equation}
i.e., if and only if 
\begin{equation}
\rho_{A+B} = (Tr_A \; \rho_{A+B})\otimes(Tr_B\;\rho_{A+B}).  
\end{equation}
Otherwise, $A$ and $B$ are said to be {\it correlated}. The concept of correlation is indistinctively classical or quantum. There is another concept, more subtle, which can exist only in quantum systems, and that is {\it 
(quantum) entanglement}. Two systems $A$ and $B$ are said to be {\it (quantum) unentangled} (or {\it separable}, or admitting a {\it local} description with ``hidden" variables, which is sometimes referred to as {\it local realism}) if and only if the corresponding density operator can be written as 
\begin{equation}
\rho_{A+B} = \sum_{i=1}^W p_i \;\rho_A^{(i)} \otimes \rho_B^{(i)} \;\;\;\; (p_i \ge 0\; \forall i; \;\sum_{i=1}^W p_i=1).
\end{equation}
 Otherwise, $A$ and $B$ are said to be {\it entangled} (or {\it nonseparable} or incompatible with  {\it local realism}). Clearly, if $A$ and $B$ are uncorrelated, they are unentangled; the opposite is not true. The definition of entanglement is not necessarily simple to implement, since it might be relatively easy in a specific case to exhibit the form of Eq. (3), but it can be nontrivial to prove that it {\it cannot} be presented in that form. Consequently, along the years appreciable effort has been dedicated to the establishment of general operational criteria, preferentially in the form of necessary and sufficient conditions whenever possible. The particular case where $A$ and $B$ are just two simple spins 1/2 is paradigmatic, and illustrates well the difficulties. 

The simplest basis for describing such systems is $|$$\uparrow>_A|$$\uparrow>_B$, $|$$\uparrow>_A|$$\downarrow>_B$, $|$$\downarrow>_A|$$\uparrow>_B$ and $|$$\downarrow>_A|$$\downarrow>_B$. All these states clearly are unentangled. Another popular basis (the Bell basis), convenient for a variety of experimental situations, is the {\it singlet} $|\Psi^-$$> \equiv \frac{1}{\sqrt 2}(|$$\uparrow>_A|$$\downarrow>_B$$-|$$\downarrow>_A|$$\uparrow>_B)$ and $|\Psi^+$$> \equiv \frac{1}{\sqrt 2}(|$$\uparrow>_A|$$\downarrow>_B$$+|$$\downarrow>_A|$$\uparrow>_B)$, $|\Phi^{\pm}$$> \equiv \frac{1}{\sqrt 2}(|$$\uparrow>_A|$$\uparrow>_B \pm |$$\downarrow>_A|$$\downarrow>_B)$ and $|\Psi^{\pm}$$>$. Each state of this basis is fully entangled. They satisfy
\begin{equation}
|\Phi^+$$><$$\Phi^+|+|\Phi^-$$><$$\Phi^-|+|\Psi^+$$><$$\Psi^+|+|\Psi^-$$><$$\Psi^-|=\hat1_{A+B} \equiv \hat1_A \otimes \hat1_B 
\end{equation}
with $Tr \; \hat 1_{A+B}=2\; Tr\; \hat 1_A= 2\; Tr \;\hat1_B= 4$.

We assume now that our bipartite system is in the so called Werner-Popescu \cite{werner,popescu} state, namely
\begin{equation}
\rho_{A+B}=\frac{1-x}{4}(|\Phi^+$$><$$\Phi^+|+|\Phi^-$$><$$\Phi^-|+|\Psi^+$$><$$\Psi^+|)+\frac{1+3x}{4}|\Psi^-$$><$$\Psi^-|
\end{equation}
or equivalently
\begin{equation}
\rho_{A+B}=\frac{1-x}{4}\;\hat1_{A+B}+x\;|\Psi^-$$><$$\Psi^-| \;\;\;(0 \le x \le 1),
\end{equation}
where we have used Eq. (4). For $x=1$ and $x=0$ we have the fully entangled EPR state and the fully random one respectively. The question arises: up to what value of $x$ is local realism possible? The use of the Bell inequality yields that the threshold cannot exceed $1/\sqrt{2} \simeq 0.71$. The use of the $\alpha$-entropic inequality \cite{horodecki1} yields a more severe restriction, namely that it cannot exceed  $1/\sqrt{3} \simeq 0.58$. The strongest result, i.e., the necessary and sufficient condition, was finally found  (using a partial transpose of the density matrix) by Peres \cite{peres}, and it is $x_c=1/3$. In a recent paper, Abe and Rajagopal \cite{aberajagopal} reobtained this result in an extremely elegant way. Let us briefly recall it.

Thermostatistically anomalous systems can, in some cases, be handled within a formalism which generalizes Boltzmann-Gibbs statistical mechanics. This formalism, introduced in 1988 \cite{tsallis}, has been used in many applications \cite{tsallis2} and has already received several verifications \cite{levy,logistic,rafelski,beck1,ion,beck2}. It is based on the entropic form
\begin{equation}
S_q=\frac{1- Tr \rho^q}{q-1}\;\;\;(q \in \cal {R}; \; $Tr$ \;\rho=$1$;\;$S$_1=-$Tr$\; \rho \ln \rho).
\end{equation}
This quantity is nonnegative ($\forall q$), concave (convex) for $q>0$ ($q<0$), and satisfies the property:
\begin{equation}
S_q(\rho_A \otimes \rho_B)=S_q(\rho_A)+S_q(\rho_B)+(1-q)S_q(\rho_A)S_q(\rho_B)\;.
\end{equation}
Consequently, it is {\it superextensive}, {\it extensive} and {\it subextensive} if $q<1$, $q=1$ and $q>1$ respectively. Also, it is extremal at equiprobability, i.e., $S_q(\hat 1/W)=\frac{W^{1-q}-1}{1-q}$ with $S_1(\hat1 /W) = \ln W$.

Ref. \cite{aberajagopal} defines the following {\it conditional entropy}
\begin{equation}
S_q(B|A)\equiv \frac{S_q(A+B)-S_q(A)}{1+(1-q)S_q(A)},
\end{equation}
where $S_q(A+B) \equiv S_q(\rho_{A+B})$ and $S_q(A) \equiv S_q(Tr_B\; \rho_{A+B})$. The conditional entropy $S_q(A|B)$ is defined in an analogous manner. Consequently, Eq. (9) implies
\begin{eqnarray}
S_q(A+B) &=& S_q(A)+S_q(B|A)+(1-q)S_q(A)S_q(B|A) \\
                 &=&S_q(B)+S_q(A|B)+(1-q)S_q(B)S_q(A|B).
\end{eqnarray} 
These expressions generalize \cite{abeaxiom} Eq. (8), which is recovered as the particular case where $\rho_{A+B}=\rho_A \otimes \rho_B$, hence $S_q(A|B)=S_q(A)$ and $S_q(B|A)=S_q(B)$. Also, for $q=1$, they reproduce one of the Shannon-Khinchin axioms for obtaining the BG form for the entropy, i.e., $S_1$. Finally, it can be shown \cite{aberajagopal} that, for all values of $q$, these expressions are consistent with the celebrated Bayes theorem. 

The entropies $S_q(A+B)$, $S_q(A)$ and $S_q(B)$ are necessarily nonnegative. This is also true for $S_q(A|B)$ and $S_q(B|A)$ {\it if the system is classical, but not necessarily if it is quantum}. Therefore, this property can be used as a criterion for local realism (see also \cite{cerfadami}). The conjecture is:
{\it Local realism is possible if and only if both $S_q(A|B)$ and $S_q(B|A)$ are nonnegative for all values of $q$}. Let us stress that there is no general reason why $S_q(A|B)$ and $S_q(B|A)$ should be equal, not even in the classical case. Although we have no general proof, it seems plausible that both $S_q(A|B)$ and $S_q(B|A)$ decrease monotonically with $q$. In that case, the conjecture becomes: {\it Local realism is possible if and only if both $S_{\infty}(A|B)$ and $S_{\infty}(B|A)$ are nonnegative}. 

Abe and Rajagopal have applied \cite{aberajagopal} this procedure to the Werner-Popescu state mentioned above, and have successfully recovered Peres' threshold $x_c=1/3$. 

In order to illustrate the simplicity of use of the present criterion, we shall assume the following state for the bipartite spin 1/2 system:
\begin{equation}
\rho_{A+B}=\frac{1-x}{4}\;|\Phi^+$$><$$\Phi^+|
+ \frac{1-y}{4}\;|\Phi^-$$><$$\Phi^-| + \frac{1-z}{4}\;|\Psi^+$$><$$\Psi^+| + \frac{1+x+y+z}{4}\;|\Psi^-$$><$$\Psi^-|
\end{equation}
or equivalently
\begin{equation}
\rho_{A+B}=\frac{1}{4}\;\hat1_{A+B}   -\frac{x}{4}\;|\Phi^+$$><$$\Phi^+|-\frac{y}{4}\;|\Phi^-$$><$$\Phi^-|-\frac{z}{4}\;|\Psi^+$$><$$\Psi^+|
+(x+y+z)\;|\Psi^-$$><$$\Psi^-|\;,
\end{equation}
with $x,y,z \le 1$. Eqs. (5) and (6) are reproduced in the $x=y=z$ case. The pure states $|\Phi^+$$>$, $|\Phi^-$$>$, $|\Psi^+$$>$ and $|\Psi^-$$>$ (EPR state) respectively correspond to $(x,y,z)=(-3,1,1),\;(1,-3,1),\;(1,1,-3)$ and $(1,1,1)$.

Let us now calculate $S_q(A+B)$. Eq. (12) implies
\begin{equation}
\rho_{A+B}^{\;q}=\Bigl(\frac{1-x}{4}\Bigr)^q\;|\Phi^+$$><$$\Phi^+|+\Bigl(\frac{1-y}{4}\Bigr)^q\;|\Phi^-$$><$$\Phi^-|+\Bigl(\frac{1-z}{4}\Bigr)^q\;|\Psi^+$$><$$\Psi^+|+\Bigl(\frac{1+x+y+z}{4}\Bigr)^q\;|\Psi^-$$><$$\Psi^-|
\end{equation}
hence
\begin{equation}
S_q(A+B)=\frac{1}{1-q}\Bigl[\Bigl(\frac{1-x}{4}\Bigr)^q+  \Bigl(\frac{1-y}{4}\Bigr)^q+\Bigl(\frac{1-z}{4}\Bigr)^q+\Bigl(\frac{1+x+y+z}{4}\Bigr)^q-1 \Bigr].
\end{equation}

Let us now calculate $S_q(A|B)$. We need to know $\rho_A=Tr_B\; \rho_{A+B}$, i.e., 
\begin{equation}
\rho_A=\frac{1-x}{4} Tr_B\; |\Phi^+$$><$$\Phi^+| +\frac{1-y}{4} Tr_B\; |\Phi^-$$><$$\Phi^-| +\frac{1-z}{4} Tr_B\; |\Psi^+$$><$$\Psi^+| +\frac{1+x+y+z}{4} Tr_B\; |\Psi^-$$><$$\Psi^-| 
\end{equation}
hence
\begin{equation}
\rho_A=\frac{1}{2} \;\hat1_A\;,
\end{equation}
where we have used the fact that $Tr_B\; |\Phi^+$$><$$\Phi^+|=Tr_B\; |\Phi^-$$><$$\Phi^-|=Tr_B\; |\Psi^+$$><$$\Psi^+|=Tr_B\; |\Psi^-$$><$$\Psi^-|=\frac{1}{2}\;\hat1_A$. Eq. (17) implies 
\begin{equation}
\rho_A^{\;q}=\frac{1}{2^q} \;\hat1_A\;,
\end{equation}
hence
\begin{equation}
S_q(A)=\frac{2^{1-q}-1}{1-q}\;.
\end{equation}
Substituting expressions (15) and (19) into Eq. (9) we obtain $S_q(A|B)$ as an explicit function of $(x,\;y,\;z;\;q)$ (See Figs. 1 and 2). Both 
$S_q(A+B)$ and $S_q(A|B)$ are invariant under the transformations $(x,y,z) \rightarrow (x,z,y)$, $(x,y,z) \rightarrow (-x-y-z,x,y)$, and the analogous ones. $S_q(A|B)=S_q(B|A)=0$ implies
\begin{equation}
\Bigl(\frac{1-x}{4}\Bigr)^q+  \Bigl(\frac{1-y}{4}\Bigr)^q+\Bigl(\frac{1-z}{4}\Bigr)^q+\Bigl(\frac{1+x+y+z}{4}\Bigr)^q = \frac{1}{2^{q-1}}\;.
\end{equation}
In the limit $q \rightarrow \infty$, this relation implies
\begin{equation}
x+y+z=1\;.
\end{equation}
In other words, if the present conjecture is correct, local realism is impossible in the neighborhood of the $|\Psi^{-}$$>$ (EPR state) if and only if $x+y+z>1$. If $x=y=z$ we recover $x_c=1/3$ (Peres criterion). If all the symmetries of the problem are used, we obtain Fig. 3. We see there that the physical space is a tetrahedron included in a $4 \times 4 \times 4$ cube. The vertices of the tetrahedron correspond to the four states of the Bell basis. Each of these vertices is also the outer vertex of a smaller tetrahedron, inside which no classical realism is possible. These four smaller tetrahedra delimit a octahedron surrounding the origin $x=y=z=0$ (state of full randomness). Classical realism is possible if and only if $(x,y,z)$ belongs to this octahedron. This geometry coincides with that obtained in \cite{horodeckicube} from a quite different standpoint.

Let us focus on the vicinity of the EPR state. If we observe carefully Fig. 2 we can see that all the curves such that $1<x+y+z <3$ exhibit, as functions of $q$, an inflexion point hereafter referred to as $q_I$. The inflexion point runs to infinity when we approach the plane $x+y+z=1$ from above (see Fig. 3), and runs to zero when we approach the point $x=y=z=1$, varying continuously in between. In all cases where the inflexion point exists, we notice that for $q>q_I$ the conditional entropy $S_q(A|B)=S_q(B|A)$ bends quickly towards minus infinity. Consequently, this point is an intrinsic characteristic of the quantum entanglement between the subsystems $A$ and $B$. Moreover, for convenience we can define the quantity $\eta \equiv 1/(1+q_I) \in [0,1]$, which plays a role analogous to an order parameter in standard critical phenomena. Indeed, in the whole region $0 \le x+y+z \le 1$ we have $\eta=0$ (``local realistic" phase); the region $1<x+y+z \le 3$ corresponds therefore to the ``nonlocal realistic" phase, the entanglement ``order parameter" $\eta$ reaching unity at the $(1,1,1)$ corner of the cube in Fig. 3. If we consider now the entire physical region (Fig. 3), we see that $\eta$ vanishes inside the central octahedron described above, and is different from zero inside the four tetrahedra neighboring respectively the four pure states of the Bell basis; for these states it is unity. At the present stage, this critical-phenomenon-like scenario is but a suggestive analogy. Indeed, $\eta$ (or any other convenient quantity related to $q_I$) cannot be considered as an order parameter in the thermodynamical sense unless several other properties are clearly understood, such as the symmetry which is broken if any, and the parameter thermodynamically conjugate to the order parameter (the associated susceptibility would diverge at the critical surface, i.e., the faces of the central octahedron). Further studies would be needed for better understanding the implications and degree of generality of the present scheme.

Summarizing, we have used the zero of the Abe-Rajagopal conditional entropy \cite{aberajagopal} as a criterion for local realism in a bipartite spin 1/2 system in the quite general state (12). The discussion of other systems and/or other states \cite{peres,GHZ} would certainly be enlightening. In particular this would clarify the degree of generality of the Abe-Rajagopal method for determining necessary and sufficient conditions for local realism \cite{horodecki2}. Generally speaking, the present work reinforces the now common understanding \cite{horodecki1,horodecki2,zurek2,popescu2,horodecki3,grigolini,aberajagopal2,vidiellabarranco,brukner} that the connections and analogies between quantum entanglement and thermodynamics are deep and fruitful.
 
One of us (C.T.) acknowledges warm hospitality at MIT, as well as enlightening remarks 
from S. Abe and A.K. Rajagopal. This work was supported in part by DARPA under the QUIC initiative.


$(a)$ tsallis@cbpf.br, tsallis@mit.edu

$(b)$ slloyd@mit.edu

$(c)$ baranger@ctp.mit.edu



\centerline{{\bf FIGURE CAPTIONS}}

\bigskip
\noindent
Fig. 1.~
$S_q(B|A)=S_q(A|B)$ versus $(x,y,z)$ for typical values of $q$: (a) $q=1/2$ for the solid lines, $q=2$ for the dashed lines, and $q=5$ for the dotted lines, along the directions $(x,0,0)$, $(x,x,0)$ and $(x,x,x)$ from top to bottom; (b) For $(x,y,z)$ along the edge joining $|\Phi^+$$>$ and $|\Psi^-$$>$ or, equivalently, $|\Phi^+$$>$ and $|\Phi^-$$>$ (notice the symmetry with regard to the $x=-1$ axis). In fact $S_q(B|A)$ varies in the same way along the six edges of the big tetrahedron indicated in Fig. 3.    

\bigskip
\noindent
Fig. 2.~
$S_q(B|A)=S_q(A|B)$ versus $q$, for typical values of $(x,y,z)$. 
The curve which for $q>0$ is the uppermost is given by $(2^{1-q}-1) /(1-q)$. The lowest curve is given by $-(2^{q-1}-1)/(q-1)$. Notice that six interesting nonuniform convergences occur at $q=0$, namely when (i) the $(x,0,0)$ curves approach, for $x \rightarrow 1$, the $(1,0,0)$ curve; (ii)  the $(x,x,0)$ curves approach, for $x \rightarrow 1$, the $(1,1,0)$ curve; (iii) the $(x,x,x)$ curves approach, for $x \rightarrow 1$, the $(1,1,1)$ curve; (iv) the $(1,x,0)$ curves approach, for $x \rightarrow 1$, the $(1,1,0)$ curve; (v) the $(1,x,x)$ curves approach, for $x \rightarrow 1$, the $(1,1,1)$ curve; (vi) the $(1,1,x)$ curves approach, for $x \rightarrow 1$, the $(1,1,1)$ curve. For $q<0$, all curves, excepting the $(1,1,1)$ one, have positive values and curvatures. The $(1,1,1)$ curve is everywhere negative both in value and curvature.

\bigskip
\noindent
Fig. 3.~
The physical space of the mixed state considered in the present paper is the tetrahedron determined by the four big circles. Every big circle and its three neighboring small circles determine a region (small tetrahedron) where no classical realism is possible. The four small tetrahedra delimit a central octahedron where classical realism is possible. The $x+y+z=1$ plane (dashed) generalizes the Peres criterion, and plays the role of a critical surface. The entanglement ``order parameter" $\eta \equiv 1/(1+q_I)$ is zero inside the central octahedron, and continuously increases when we approach the four vertices of the big tetrahedron, where $\eta=1$.

\vfill


\begin{thebibliography}{99}

\bibitem{EPR}A. Einstein, B. Podolsky and N. Rosen, Phys. Rev. {\bf 47}, 777 (1935).

\bibitem{schroedinger}E. Schroedinger, Proc. Cambridge Philos. Soc. {\bf 31}, 555 (1935).

\bibitem{ekert}A. Ekert, Phys. Rev. Lett. {\bf 67}, 661 (1991).

\bibitem{bennett}C. Bennett, G. Brassard, C. Crepeau, R. Jozsa, A. Peres and W.K. Wootters, Phys. Rev. Lett. {\bf 70}, 1895 (1993).

\bibitem{barenco}A. Barenco, D. Deutsch, A. Ekert and R. Jozsa, Phys. Rev. Lett. {\bf 74}, 4083 (1995).

\bibitem{zurek}R. Laflamme, C. Miquel, J.P. Paz and W.H. Zurek, Phys.  Rev. Lett. {\bf 77}, 198 (1996).

\bibitem{lloyd}H. Touchette and S. Lloyd, Phys. Rev. Lett. {\bf 84}, 1156 (2000) 

\bibitem{quantumchaos}A. Peres, {\it Quantum theory: Concepts and methods}, (Kluwer, Dordrecht, 1993).

\bibitem{werner}R.F. Werner, Phys. Rev. A {\bf 40}, 4277 (1989).

\bibitem{popescu}S. Popescu, Phys. Rev. Lett. {\bf 72}, 797 (1994).

\bibitem{horodecki1}M. Horodecki, P. Horodecki and R. Horodecki, Phys. Lett. A {\bf 210}, 377 (1996).

\bibitem{peres}A. Peres, Phys. Rev. Lett. {\bf 77}, 1413 (1996).

\bibitem{aberajagopal}S. Abe and A.K. Rajagopal, Physica A (2000), in press [quant-ph/0001085].

\bibitem{tsallis}C. Tsallis, J. Stat. Phys. {\bf 52}, 479 (1988);
E.M.F. Curado and C. Tsallis, J. Phys. A {\bf 24}, L69 (1991) [Corrigenda: {\bf
24}, 3187 (1991) and {\bf 25}, 1019 (1992)]; C. Tsallis, R.S. Mendes and
A.R. Plastino, Physica A {\bf 261}, 534 (1998). A regularly updated
bibliography on the subject is accessible at
http://tsallis.cat.cbpf.br/biblio.htm

\bibitem{tsallis2}
C. Tsallis, in {\it Nonextensive Statistical 
Mechanics and Thermodynamics}, eds. S.R.A. Salinas and C. Tsallis, 
Braz. J. Phys. {\bf 29}, 1 (1999) 
[http://sbf.if.usp.br/ WWW$_{-}$pages/Journals/BJP/Vol29/Num1/index.htm]; 
C. Tsallis, in {\it Nonextensive Statistical Mechanics and Its Applications}, 
eds. S. Abe and Y. Okamoto, Series {\it Lecture Notes in Physics} (Springer-Verlag, Berlin, 2000), in press.

\bibitem{levy}P.A. Alemany and D.H. Zanette, Phys. Rev. E {\bf 49}, R956 (1994); D.H. Zanette and P.A. Alemany, Phys. Rev. Lett. {\bf 75}, 366 (1995); C. Tsallis, S.V.F Levy, A.M.C. de Souza and R. Maynard, Phys. Rev. Lett. {\bf 75}, 3589 (1995); Erratum: Phys. Rev. Lett. {\bf 77}, 5442 (1996); M. Buiatti, P. Grigolini and A. Montagnini, Phys. Rev. Lett. {\bf 82}, 3383 (1999); D. Prato and C. Tsallis, Phys. Rev. E {\bf 60}, 2398 (1999); C. Budde, D. Prato and M. Re, preprint (2000) [cond-mat/0007038]; A. Robledo, Phys. Rev. Lett. {\bf 83}, 2289  (1999).

\bibitem{logistic}
C. Tsallis, A.R. Plastino and W.-M. Zheng, Chaos, 
Solitons and Fractals {\bf 8}, 885 (1997); 
U.M.S. Costa, M.L. Lyra, A.R. Plastino and C. Tsallis, 
Phys. Rev. E {\bf 56}, 245 (1997);
M.L. Lyra and C. Tsallis, Phys. Rev. Lett. {\bf 80}, 53 (1998);
U. Tirnakli, C. Tsallis and M.L. Lyra, Eur. Phys. J. B. {\bf 10}, 
309 (1999); 
C.R. da Silva, H.R. da Cruz and M.L. Lyra, Braz. J. Phys. {\bf 29}, 144 (1999) 
[http://sbf.if.usp.br/WWW$_{-}$pages/Journals/BJP/Vol29/Num1/index.htm]; V. Latora, M. Baranger, A. Rapisarda and C. Tsallis, Phys. Lett. A (2000), in press [cond-mat/9907412].

\bibitem{rafelski}D.B. Walton and J. Rafelski, Phys. Rev. Lett. {\bf 84}, 31 (2000).

\bibitem{beck1}C. Beck, Physica A {\bf 277}, 115 (2000); T. Arimitsu and N. Arimitsu, Phys. Rev. E {\bf 61}, 3237 (2000) and J. Phys. A {\bf 33}, L235 (2000).

\bibitem{ion}M.L.D. Ion and D.B. Ion, Phys. Lett. B {\bf 482}, 57 (2000). 

\bibitem{beck2}I. Bediaga, E.M.F. Curado and J. Miranda, Physica A (2000), in press [hep-ph/9905255]; Beck, Physica A (2000), in press [hep-ph/0004225].

\bibitem{abeaxiom}S. Abe, Phys. Lett. A {\bf 271}, 74 (2000).

\bibitem{cerfadami}N.J. Cerf and C. Adami, Phys. Rev. Lett. {\bf 79}, 5194 (1997); Phys. Rev. A {\bf 60}, 893 (1999); A.K. Rajagopal, in {\it Nonextensive Statistical Mechanics and its Applications}, eds. S. Abe and Y. Okamoto,  {\it Lecture Notes in Physics} (Springer-Verlag, Berlin, 2000), in press.

\bibitem{horodeckicube}M. Horodecki, P. Horodecki and R. Horodecki, Phys. Rev. Lett. {\bf 78}, 574 (1997). 

\bibitem{GHZ}D.M. Greenberger, M. A. Horne and A. Zeilinger, in {\it Bell's theorem, Quantum Theory , and Conceptions of the Universe}, ed. M. Kafatos (Kluwer Academic, Dordrecht, 1989), p. 73;  D.M. Greenberger, M. A. Horne, A. Shimony and A. Zeilinger, Am. J. Phys. {\bf 58}, 1131 (1990).

\bibitem{horodecki2}M. Horodecki, P. Horodecki and R. Horodecki, Phys. Lett. A {\bf 223}, 1 (1996).

\bibitem{zurek2}W.H. Zurek, S. Habib and J.P. Paz, Phys. Rev. Lett. {\bf 70}, 1187 (1993).

\bibitem{popescu2}S. Popescu and D. Rohrlich, Phys. Rev. A {\bf 56}, R3319 (1997).

\bibitem{horodecki3}M. Horodecki, P. Horodecki and R. Horodecki, Phys. Rev. Lett. {\bf 80}, 5239 (1998).

\bibitem{grigolini}D. Vitali and P. Grigolini, Phys. Lett. A {\bf 249}, 248 (1998).

\bibitem{aberajagopal2}S. Abe and A.K. Rajagopal, Phys. Rev. A {\bf 60}, 3461 (1999).

\bibitem{vidiellabarranco}A. Vidiella-Barranco, Phys. Lett. A {\bf 260}, 335 (1999). 

\bibitem{brukner}C. Brukner and A. Zeilinger, {\it Conceptual Inadequacy of the Shannon Information in Quantum Measurements}, preprint (2000) [quant-ph/0006087]. 
     

\end{thebibliography}
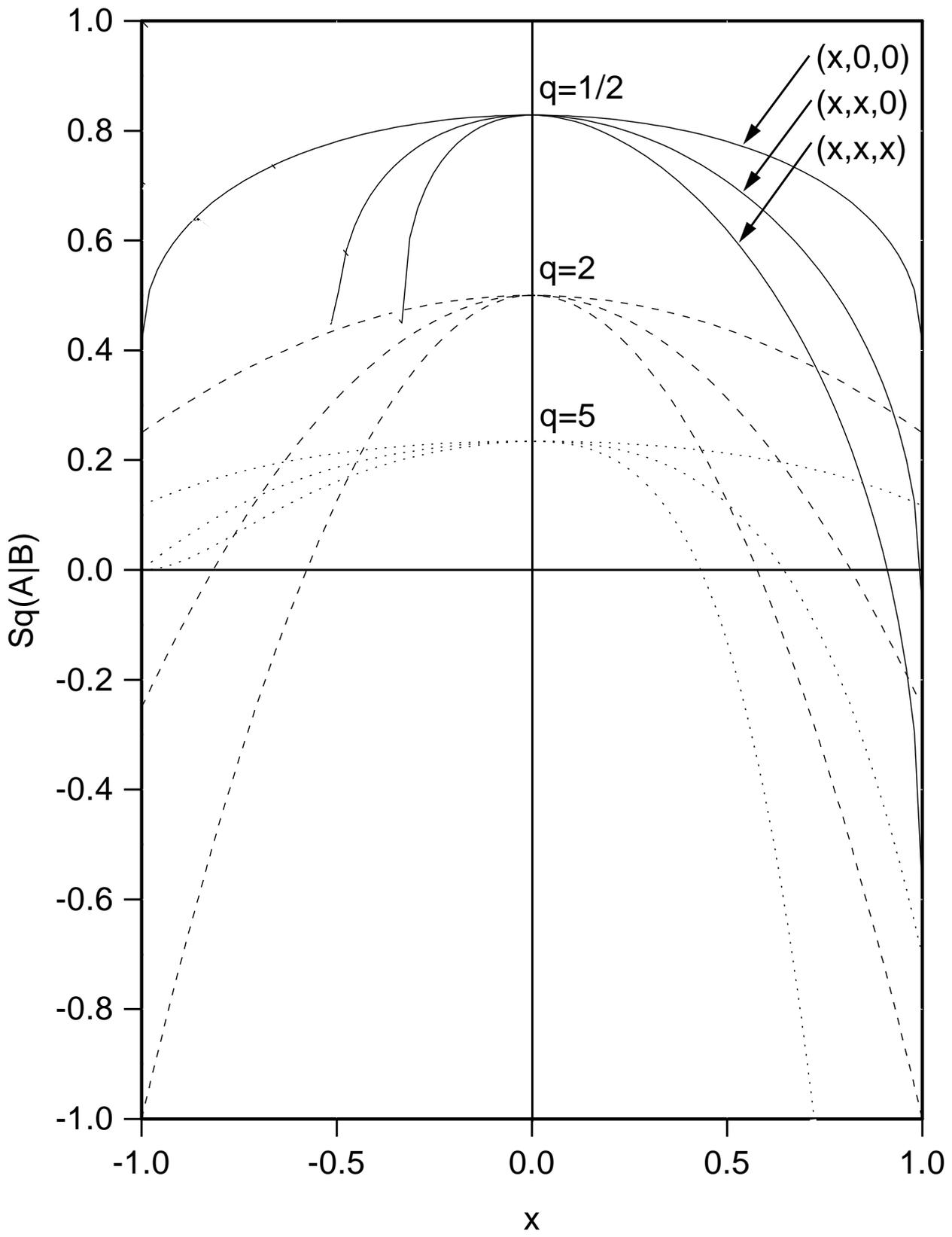
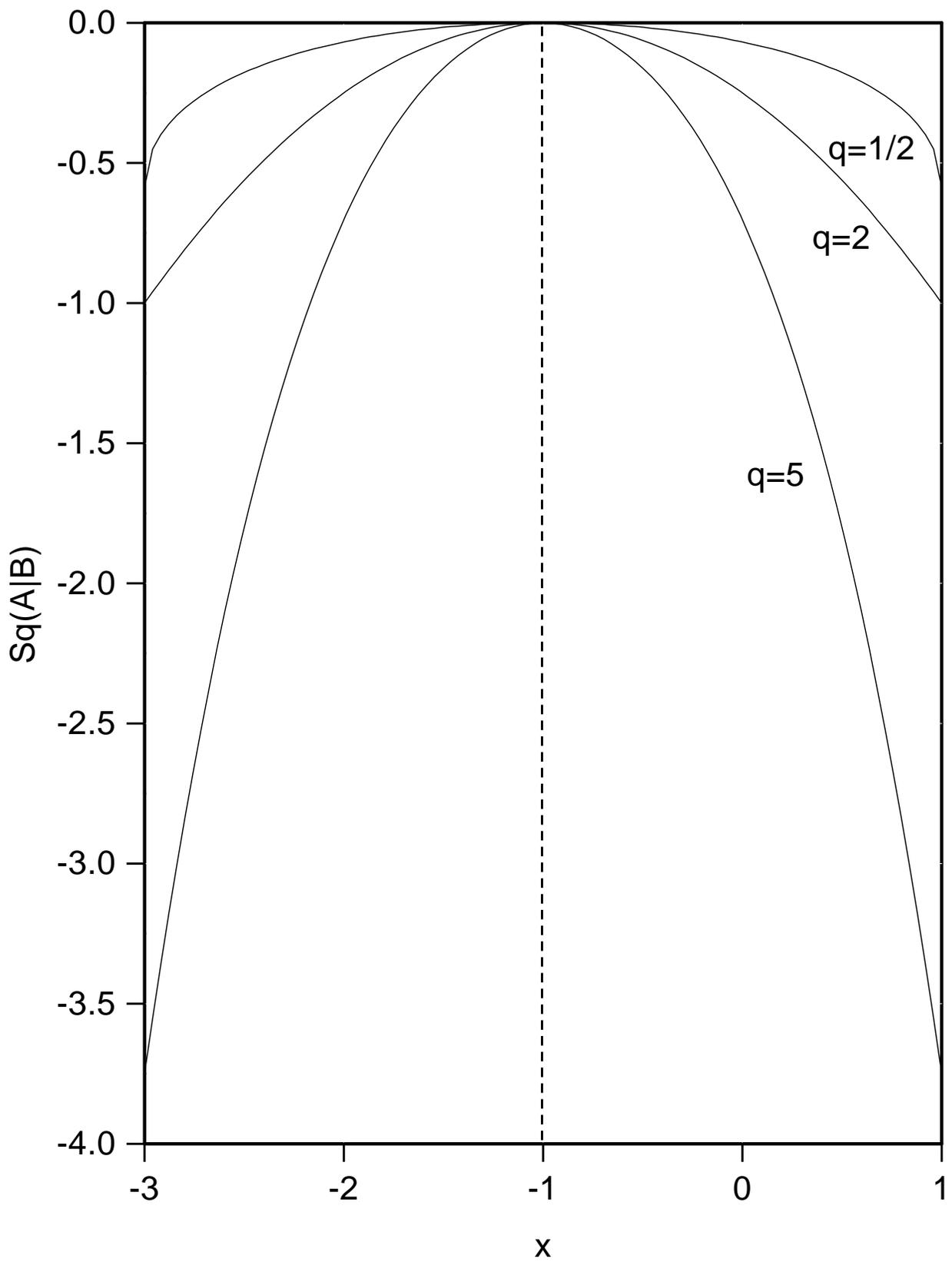
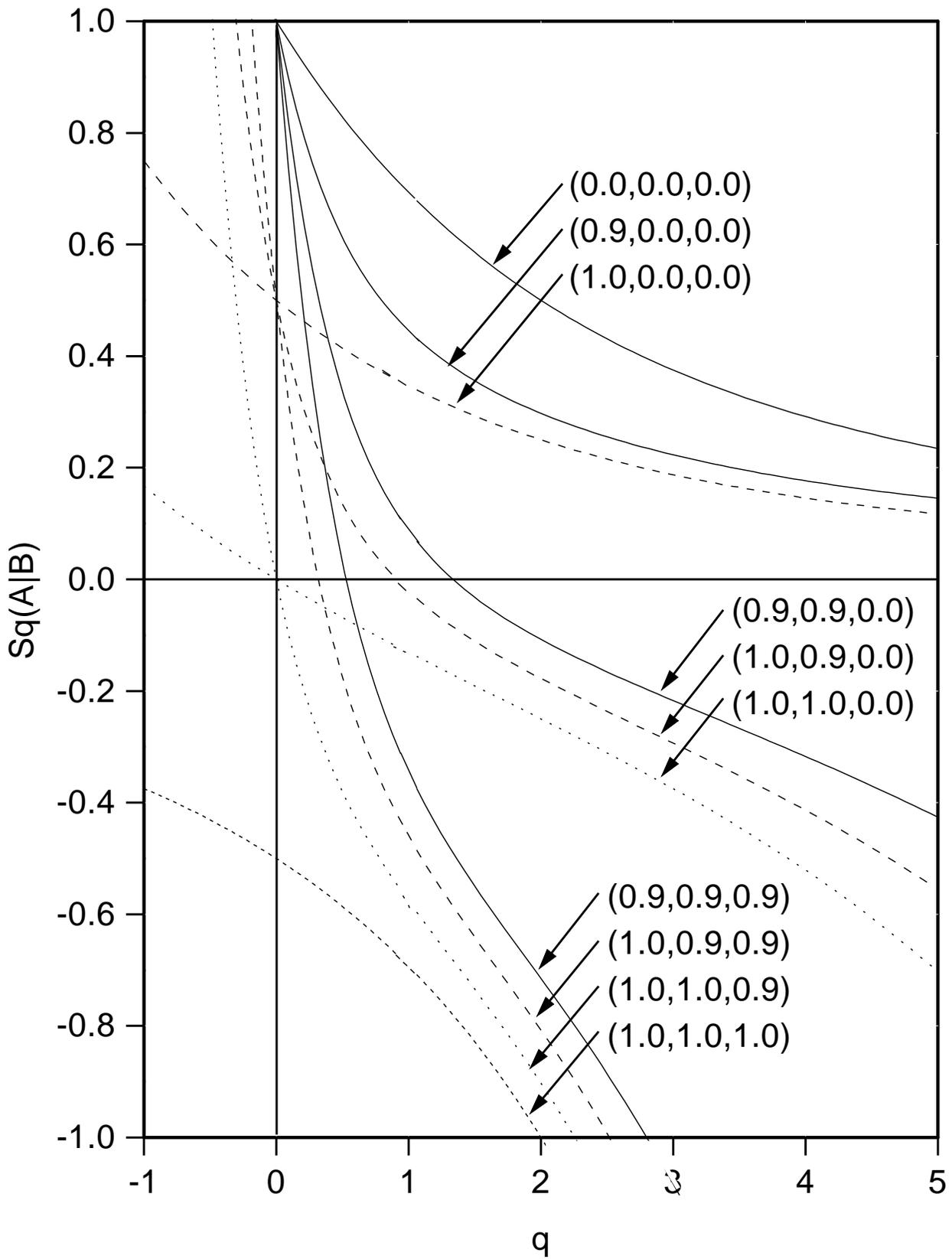
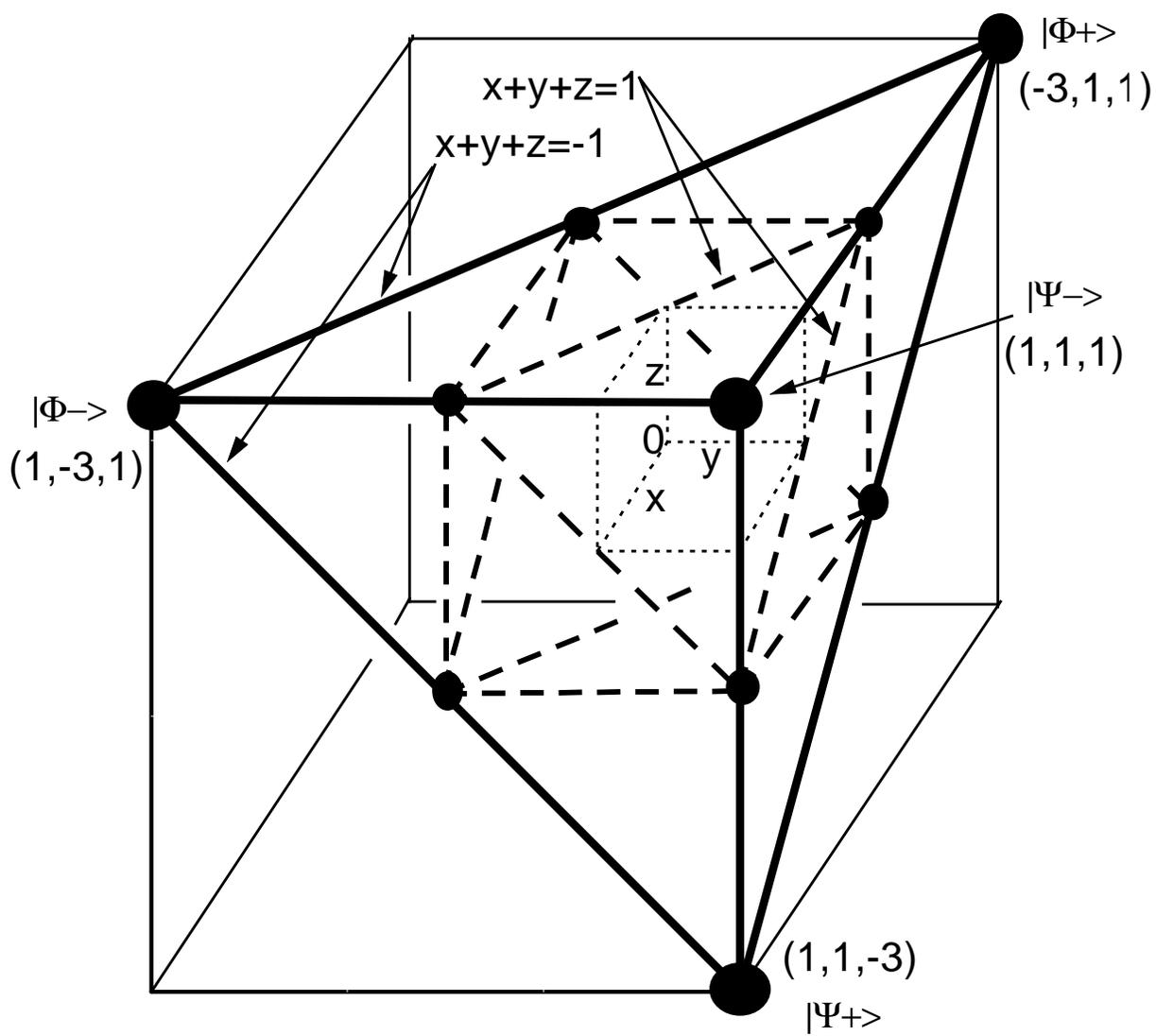
\end{document}